# Topological changes at the jamming and gel transition of a reversible polymeric network

, Joris Billen[1], Mark Wilson[1], Avinoam Rabinovitch[2], and Arlette R.C. Baljon[1]
[1]Department of Physics, San Diego State University, San Diego, CA 92128, USA
[2]Department of Physics, Ben-Gurion University of the Negev, Beer-Sheva, 84105, Israel

We investigate the network topologies of an ensemble of telechelic polymers. The telechelic polymers serve as "links" between "nodes", which consist of aggregates of their associating endgroups. Our analysis shows that the degree distribution of the systems is bimodal and consists of two Poissonian distributions with different average degrees. The number of nodes in each of them as well as the distribution of links depends on temperature. By comparing the eigenvalue spectra of the simulated network gels with those of reconstructed networks, the most likely topology at each temperature is determined. Topological changes occur at the jamming and gel transition temperatures reported in our previous studies [1]. The jammed state topology can be described by a robust bimodal network in which hubs or superpeer nodes are linked among each other and peer nodes are linked to superpeers.

Reversible polymeric gels contain networks of physically associating polymers. These are copolymers incorporating soluble fragments along with a small fraction of insoluble chemical groups or linkers, which strongly attract each other. Different chemical units may be used as linkers depending on the solvent, e.g., hydrophobic fragments on water-soluble polymers in aqueous solutions, and ionic groups on ionomers in organic solvents. The linkers form stable aggregates, which serve as temporary junctions in the resulting network structure [2,3]. A transition from a fluid "sol" state to a glassy "gel" state is obtained by increasing the polymer concentration or decreasing the temperature. Transitions can also be triggered by applying a shear stress. It has been argued that the slowdown in the dynamics of the system when approaching the gel transition is not exclusively due to the longer lifetime of the aggregates. Structural changes in the network are believed to be a factor too [4,5]. In previous work [1] we have indeed shown that it is possible to define a jamming transition for gels similar to that defined for glassy systems, granular matter, and other amorphous systems. Such a jamming transition is believed to result from a self-organization of the system [6].

In this letter, we use graph theory to quantify the topology of a simulated gel network (SGN). Details of the simulations can be found in [1]. We will show that the topology changes as a function of temperature and point out differences above and below the jamming transition. All simulations are carried out on a system of 1000 telechelic polymer chains. Each polymer contained 8 beads. Both end groups are linkers. Non-bonded interactions between the beads are modeled using a purely repulsive Lennard-Jones (LJ) potential. All quantities are expressed in terms of the parameters ($\sigma,\varepsilon$) of this potential. Beads connected by the chain structure interact through a FENE potential. Beads at chain ends can form junctions. They are modeled by a FENE potential with the same parameters. The positions of the beads are updated in Molecular Dynamics simulations using the aforementioned force-fields. Attempts to create and destroy junctions are Monte Carlo moves [7]. The probability of success depends on $\exp(-\Delta U/kT)$, where $\Delta U$ is the difference in energy between the old and new state. It equals the potential energy of a junction and hence is the sum of the FENE potential and a negative constant $U_{assoc}= -22$ that models a dissociation barrier. The temperature is controlled by coupling the simulation cell to a heat bath. At each temperature the system is equilibrated for at least $5000\tau$ before structural properties are sampled. Data at the lower temperatures are all obtained starting from a well equilibrated configuration at T=1.5. The system is then slowly cooled at a rate of $2.500\tau$ per $\Delta T=0.1$, in order to obtain the desired temperatures. Data obtained by cooling the system from different initial states at T=1.5 are identical within statistical fluctuations. Moreover the statistical averages of structural properties do not change appreciably over time, even though the system shows rich dynamics with individual aggregates dissolving and new ones forming. Aggregates of end groups form and their sizes increase as temperature decreases. At low temperatures the system undergoes a gel transition: due to the aggregates an extended network forms that prohibits flow. Since the density is approximately 30%, no sign of a glass transition is observed.

Our previous study [1] has defined four characteristic temperatures for the gel transition. At low temperature the relaxation time as a function of temperature diverges either as a stretched exponential at $T_0=0.29$ or as a power-law at $T_c=0.4$. Above $T_A=0.75$ the dependence of relaxation time on temperature becomes Arrhenius. At the micelle transition temperature $T_m=0.51$ the number of reversible bonds strongly increases and the specific heat peaks. Below $T_m$ the overall structure of the reversible network changes and a peak in the micelle size distribution becomes visible [1,8]. This state resembles self-organized jammed states in other amorphous systems [6,9,10]. Between $T_m$ and $T_0$ collective modes of relaxation are still available to the system and cause a net flow over long time periods. In this letter we investigate the structure of the SGN and how it changes with temperature. To this end, we compare it with that of complex networks found elsewhere in nature [11]. A complex network can be described as a set of nodes with links in between. In the well-known Erdos-Renyi (ER) random network, every pair of nodes is linked with a probability p. The degree distribution of this network, which describes the number of links k per node, is Poissonian:

$$P(k) = \frac{\langle k \rangle^k e^{-\langle k \rangle}}{k!} \qquad (1)$$

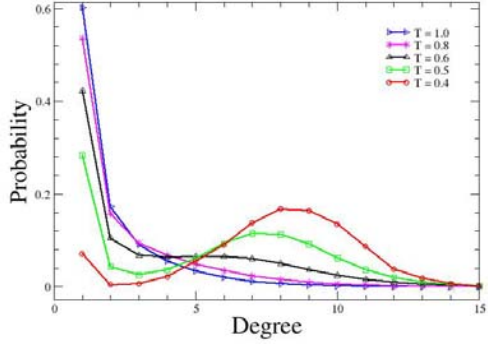

FIG. 1: Degree distributions of the networks obtained from the simulation data at various temperatures.

The average degree $\langle k \rangle = Np = \frac{2l}{N}$, where $l$ equals the number of links and N the number of nodes. In order to compare the simulated gels to other complex networks, we call an aggregate of chain end groups a node. A polymer chain of which both ends belong to different aggregates is identified as a link. If more than one polymer links the same aggregates, this still counts as one link. The degree distribution of the networks defined this way is shown in Fig 1. At the lower temperatures a peak at finite k is observed, yet the distribution cannot be described by a single Poissonian. As shown in Fig 2 for T=0.55, a superposition of two Poisson distributions with different values of <k> does fit the data.

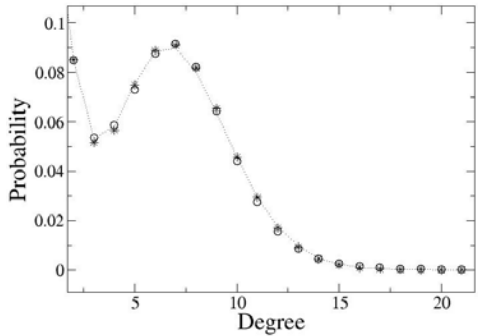

FIG 2: Degree distribution for T=0.55 (circles). The dotted line shows a fit to Eq. 2 using the values in Table 1. The stars result from a reconstruction of the gel network as described in the text.

$$P(k) = N_S \frac{\langle k \rangle_S^k e^{-\langle k \rangle_S}}{k!} + N_P \frac{\langle k \rangle_P^k e^{-\langle k \rangle_P}}{k!} \qquad (2)$$

Such a bimodal degree distribution is characteristic of complex networks which contain two types of nodes. Nodes in the distribution with the higher <k> are called hubs or "superpeers" (S), those in the other distribution "peers" (P) [12]. $N_S$ and $N_P$ are the fraction of superpeer and peer nodes. The probability to form links between two superpeers ($p_{SS}$) differs from that to form links between peers and superpeers ($p_{PS}$) or between peers ($p_{PP}$). This results in average degree of superpeer nodes <k>$_S$ and peer nodes <k>$_P$ given by:

$$\langle k \rangle_S = p_{SS} N_S + p_{PS} N_P$$
$$\langle k \rangle_P = p_{PS} N_S + p_{PP} N_P \qquad (4)$$

In table 1 we list the values of $N_S$, $N_P$, <k>$_S$, and <k>$_P$ of the fits to the SGN for a range of temperatures. With decreasing temperature $N_S$ increases. Below T=0.4 $N_P$=0 and a single distribution of superpeers remains. The correlation coefficient (CC) of the fit is excellent at high temperatures, but becomes less accurate at temperatures below T=0.5. We found that the single distribution below T=0.4 has a slightly higher variance than predicted by a Poissonian model.

| Temperature | $N_S$ | $N_P$ | <k>$_S$ | <k>$_P$ | CC |
|---|---|---|---|---|---|
| 5.0 | 0.063 | 0.937 | 1.512 | 0.420 | 1.000 |
| 3.5 | 0.087 | 0.913 | 1.796 | 0.499 | 1.000 |
| 3.0 | 0.100 | 0.900 | 1.915 | 0.533 | 1.000 |
| 2.6 | 0.110 | 0.890 | 2.038 | 0.553 | 1.000 |
| 2.2 | 0.124 | 0.876 | 2.238 | 0.621 | 1.000 |
| 1.8 | 0.137 | 0.863 | 2.551 | 0.716 | 1.000 |
| 1.5 | 0.165 | 0.835 | 2.731 | 0.735 | 1.000 |
| 1.2 | 0.193 | 0.807 | 3.264 | 0.890 | 1.000 |
| 1.0 | 0.226 | 0.774 | 3.867 | 1.045 | 1.000 |
| 0.8 | 0.303 | 0.697 | 4.784 | 1.266 | 1.000 |
| 0.7 | 0.380 | 0.620 | 5.544 | 1.402 | 0.999 |
| 0.6 | 0.525 | 0.475 | 6.552 | 1.387 | 1.000 |
| 0.55 | 0.604 | 0.396 | 7.131 | 0.997 | 1.000 |
| 0.525 | 0.687 | 0.313 | 7.461 | 0.943 | 0.997 |
| 0.5 | 0.777 | 0.223 | 7.765 | 0.830 | 0.991 |
| 0.45 | 0.991 | 0.009 | 8.356 | 0.984 | 0.976 |
| 0.4 | 1.0 | 0.0 | 8.620 | | 0.976 |
| 0.35 | 1.0 | 0.0 | 8.732 | | 0.946 |
| 0.3 | 1.0 | 0.0 | 8.720 | | 0.960 |

TABLE 1: Results of fits of degree distributions to Eq. 3

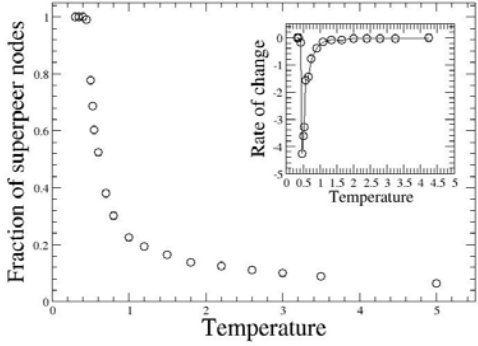

FIG 3: Fraction of superpeers as a function of temperature. The inset shows the rate of change of the fraction of superpeers.

Nevertheless the data indicate that the SGN can be described by two Poissonians, whose relative contribution depends on temperature. This is shown in Fig 3, where the fraction of superpeer nodes $N_S$ is plotted. The inset shows its rate of change. This rate peaks around T=0.5. Hence the micelle transition is not just characterized by an increase in aggregate sizes, but also by a change in the structural organization of the network formed by these aggregates.

Besides the degree distribution, the clustering coefficient is an important measure of network topology. For a particular node, it is defined as the fraction of its neighbors that connect among themselves. The clustering coefficient of the entire network is obtained by averaging over all nodes of degree two and up. Fig. 4 shows the clustering coefficient as a function of temperature for the SGN. The values are much larger than those of a random (ER) network, which can be calculated from:

$$C = \frac{\langle k \rangle}{N}\left[\frac{\langle k^2 \rangle - \langle k \rangle^2}{\langle k \rangle^2}\right]^2 \qquad (4)$$

,where N is the total number of nodes.

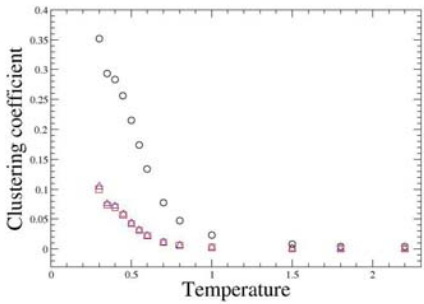

FIG 4: The clustering coefficient as a function of temperature. The circles are for the SGN. After rewiring the values change to those indicated by the triangles. The squares are the calculated values from Eq 4.

This is due to spatial effects. In the simulations, the maximum size of a link between nodes is dictated by the length of the chain molecules. In order to test this assumption, we rewired [14] the network. To this end, starting with a configuration taken from the simulations, we switched the endpoints of two links. Such a switch preserves the degree distribution. After many switches a new network is obtained similar to the old one, except that now links of all lengths are allowed. It is easy to see, that during this process the clustering coefficient decreases. Its steady state values are shown in Fig 4. As one can see the clustering coefficients of the rewired networks are very close to those predicted by Eq. 4.

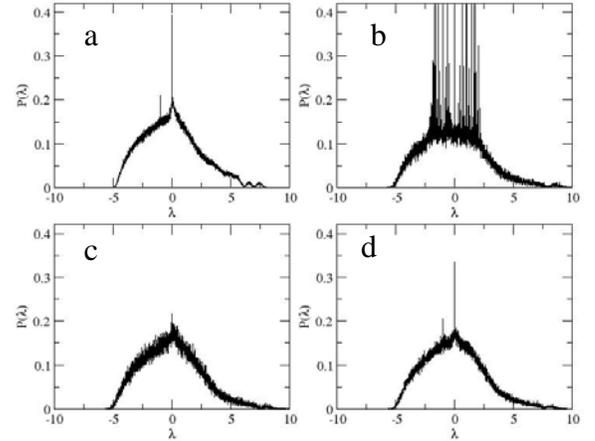

FIG 5: A comparison of the spectral density of reconstructed bimodal graphs with that of the SGN (a). In (b) the networks of peers and superpeers are independent ($p_{PS}=0$), (c) is for the case that peers do not link among themselves ($p_{PP}=0$), and (d) shows the best match to the SGN, which is obtained for $p_{SS}=0.04$, $p_{PP}=0.002$, $p_{PS}=0.009$.

Knowledge of these clustering coefficients allows us to reconstruct artificial networks representing the SGN at each temperature from the data given in Table 1. The reconstruction is done with the measured values of $N_S$, $N_P$, $\langle k \rangle_S$ and $\langle k \rangle_P$. First, the desired number of superpeers and peers are randomly placed in a spatial cell. Next, links are added in such a way that the desired values of $\langle k \rangle_S$ and $\langle k \rangle_P$ are obtained. During this process only links shorter than a certain cutoff distance are allowed. The cutoff length is adjusted such that the clustering coefficient is equal to that measured for the simulated gel network. The resulting degree distribution for T=0.55 is shown in Fig 2. We found that the cutoff distance yielding the best fit equals 6.6 $\sigma$.

There is still one degree of freedom in choosing the three probabilities in such a way that Eq. 3 is satisfied. The degree distribution is insensitive to the exact choice of the probabilities. Hence another property is needed to decide which values of the probabilities mimic the simulated gel networks best. To this end, we calculate the spectral density [15] for the SGN and compare it with that of reconstructed networks for a range of probabilities. The results for T=0.55 are shown in Fig. 5. When $p_{PS}$=0 (b), many peaks in the network indicate the abundance of disconnected clusters. A structure in which all peers are only connected to superpeers (c) is the smoothest. The best match is obtained at $p_{SS}$=0.04, $p_{PP}$=0.002, $p_{PS}$=0.009 (d). In particular, the peak in the spectrum at $\lambda$= -1 matches up well. We found that the fraction of the "giant network component" [11] in the simulated gel and reconstructed network at these settings are comparable as well.

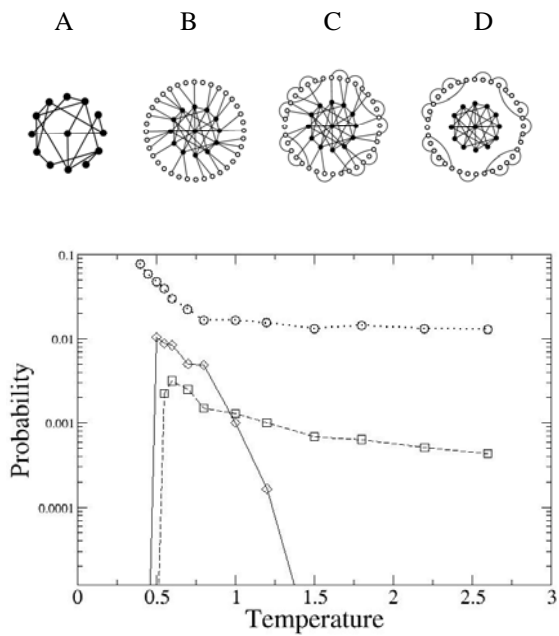

FIG 6: Probabilities to form links as a function of temperature. Circles for $p_{SS}$, squares for $p_{PP}$, and diamonds for $p_{PS}$. Lines are guides for the eye. The cartoons show the transitions in the peer-superpeer networks. Superpeers are shown as closed dots and peers as open dots.

A similar comparison is performed for all temperatures. The results are shown in Fig 6. Sketches of the peer-superpeer network are displayed as well. At the highest temperatures the system consists of two separate networks (D). If the temperature is lowered, links between peers and superpeers are formed (C). Gradually the number of peers and the number of links in between them decrease. At T=0.5 the number of links between peers vanishes and every peer is linked to a superpeer (B). Finally below T=0.4 only superpeers and links between them remain (A).

We conclude that during the transition from a fluid to a gel state the SGN undergoes several topological changes. These were studied by means of graph theory, considering aggregates as nodes and polymer chains as links. We showed that, after rewiring links to eliminate spatial effects, the SGN has the properties of a bimodal random graph. Reconstruction of representative bimodal graphs has allowed us to investigate their topology in detail. Spatial effects were accounted for by restricting the size of links in such a way that the clustering coefficient of the reconstructed graphs matched that of the SGN. By comparing spectral densities we were able to determine the number of links between peers and between superpeers, as well as those connecting peers to superpeers. We detected that significant changes in the topology occur at temperatures that in previous work [1] we have characterized as transition temperatures. Most importantly, below $T_m$ links between the regular nodes disappear. The resulting topology has been widely studied in the literature [16] and was shown to be extremely robust. These robust networks are less likely to flow due to thermal fluctuations of aggregates than those at higher temperature. Finally, below $T_c$ peers disappear and a single degree distribution remains.

The authors gratefully acknowledge support by a grant from the NSF under Grant No. DMR0517201.